\documentclass[letter]{aa} 
\usepackage{graphicx}
\usepackage{txfonts}
\usepackage{array}
\usepackage{natbib}
\usepackage{longtable}
\usepackage{float}

\newcommand{\secp}{\mbox{\rlap{.}$''$}}
\newcommand{\secsecp}{\mbox{\rlap{.}$^s$}}

\begin{document}

\title{Searching for Trans Ethyl Methyl Ether in Orion KL
\thanks{This paper makes use of the following ALMA data: ADS/JAO.ALMA$\#$2011.0.00009.SV.
ALMA is a partnership of ESO (representing its member states), NSF (USA), and NINS (Japan) with NRC (Canada), NSC, and ASIAA (Taiwan), and KASI (Republic of Korea),
in cooperation with the Republic of Chile. The Joint ALMA Observatory is operated by ESO,
AUI/NRAO, and NAOJ. This work was also based on observations carried out with the 
IRAM 30-meter telescope. IRAM is supported by INSU/CNRS (France), 
MPG (Germany), and IGN (Spain).}\fnmsep\thanks{Appendix A (online Figures and Tables) is only available in
     electronic form via http://www.edpscience.org}}


  \author{B. Tercero\inst{1}
        \and J. Cernicharo\inst{1}
        \and A. L\'opez\inst{1,2}
        \and N. Brouillet\inst{3}
        \and L. Kolesnikov\'a\inst{4}
        \and R. A. Motiyenko\inst{5}
        \and L. Margul\`es\inst{5}
        \and J. L. Alonso\inst{4}
  \and J.-C. Guillemin\inst{6}
}

\institute{Grupo de Astrof\'isica Molecular. Instituto de CC. de Materiales de Madrid (ICMM-CSIC).
Sor Juana In\'es de la Cruz, 3, Cantoblanco, 28049 Madrid, Spain.
\and Dpto. de Astrof\'isica, CAB. INTA-CSIC. Crta Torrej\'on-Ajalvir, km. 4. 28850 Torrej\'on
de Ardoz. Madrid. Spain.
\and Univ. Bordeaux, LAB, UMR 5804, F-33270 Floirac, France; CNRS, LAB, UMR 5804, F-33270 Floirac, France.
\and Grupo de Espectroscop\'ia Molecular (GEM), Edificio Quifima, \'Area de Qu\'imica-F\'isica, 
Laboratorios de Espectroscop\'ia y Bioespectroscop\'ia, Parque Cient\'ifico UVa, 
Unidad Asociada CSIC, Universidad de Valladolid, 47011 Valladolid, Spain.
\and Laboratoire de Physique des Lasers, Atomes, et Mol\'ecules, UMR CNRS 8523, 
Universit\'e de Lille I, F-59655 Villeneuve d'Ascq C\'edex, France.
\and Institut des Sciences Chimiques de Rennes, Ecole Nationale Sup\'erieure de Chimie de Rennes,
CNRS, UMR 6226, 11 All\'ee de Beaulieu, CS 50837, 35708 Rennes Cedex 7, France.\\
\email{b.tercero@icmm.csic.es; jose.cernicharo@csic.es}
}
   \date{Received ...; accepted ...}

\abstract{We report on the tentative detection of $trans$ Ethyl Methyl Ether (tEME), t-CH$_3$CH$_2$OCH$_3$, 
through the identification of a large number of rotational lines 
from each one of the spin states of the molecule 
towards Orion KL.
We also search for $gauche$-$trans$-n-propanol, Gt-n-CH$_3$CH$_2$CH$_2$OH, an isomer of tEME
in the same source.
We have identified lines of both species  in the IRAM 30m line survey and in the ALMA Science Verification data.
We have obtained ALMA maps   to establish the spatial distribution of these species. Whereas tEME
mainly arises from the compact ridge component of Orion, Gt-n-propanol appears at the emission peak of ethanol
(south hot core).
The derived column densities of these species at the location of their emission peaks
are $\leq$(4.0$\pm$0.8)$\times$10$^{15}$\,cm$^{-2}$ and $\leq$(1.0$\pm$0.2)$\times$10$^{15}$\,cm$^{-2}$
for tEME and Gt-n-propanol, respectively. The rotational temperature is $\sim$100\,K for both molecules.
We also provide maps of CH$_3$OCOH, CH$_3$CH$_2$OCOH,
CH$_3$OCH$_3$, CH$_3$OH, and CH$_3$CH$_2$OH  to compare the
distribution of these organic saturated O-bearing species containing methyl and ethyl groups in this region.
Abundance ratios of related species and upper limits to the abundances of
non-detected ethers are provided. We derive an abundance ratio $N$(CH$_3$OCH$_3$)/$N$(tEME)\,$\geq$\,150
in the compact ridge of Orion.}


\titlerunning{Trans Ethyl Methyl Ether in space} 
\authorrunning{B. Tercero et al.}

\maketitle

\section{Introduction}

The spectral millimeter-wave survey of Orion KL carried
out with the IRAM 30m radio telescope \citep{Tercero2010, Tercero2012} shows
more than 15400 spectral features of which about 11000 have been identified
and attributed to 50 molecules (199 different isotopologues and vibrational modes).
To date, there have been several works based on these data.
As the result of a fruitful
collaboration with spectroscopy laboratories, 3000 previously unidentified lines 
have been assigned to new species in the
interstellar medium (ISM).
We have detected in
space 16 new isotopologues and vibrationally excited states of 
abundant molecules in Orion for the first time \citep{Demyk2007, Margules2009, Margules2010, Carvajal2009,
Tercero2012, Motiyenko2012, Daly2013, Coudert2013, Haykal2014, Lopez2014} as well as four
new molecules 
\citep{Tercero2013, Cernicharo2013, Kolesnikova2014}.
These identifications 
reduce the number of unidentified lines and mitigate line confusion in the spectra.
Nevertheless, many features still remain  unidentified and
correspond  to new species that we have to search and identify.
Formates, ethers, acetates, alcohols, and cyanides are the best candidates for
this purpose in Orion.

The recent search for $trans$ Ethyl Methyl Ether (tEME)
in selected hot cores (Sgr B2(N-LMH) and W51 e1/e2) by \citet{Carroll2015}
only provides  upper limits to tEME. Hence, 
the results from that work do not confirm the previous tentative identification of this species by
\citet{Fuchs2005} towards W51 e1/e2.

A systematic line survey with most weeds removed permits us to
address the problem of the abundances of isomers and derivatives
of key species, such as methyl formate (A. L\'opez et al. in preparation), through
combined IRAM and ALMA studies.

In this Letter, we report on the tentative detection of tEME towards
the compact ridge (CR) of Orion KL. We have detected emission
of features arising from the five spin states 
at 3, 2, and 1\,mm with the IRAM 30m telescope
and the ALMA interferometer. In addition, several unidentified lines of these
data have been identified as belonging to the $gauche$-$trans$ conformer of
n-propanol (an isomer of tEME).
ALMA maps of organic saturated O-bearing species containing methyl, ethyl, and propyl groups,
abundance ratios of related species, and upper limits to the column
densities of non-detected ethers are presented and discussed in Sect.\,\ref{sect_dis}.

\section{Observations}
\label{sect_obs}
\textbf{IRAM 30m:} New data of the IRAM 30m telescope, which complement and improve those of \citet{Tercero2010},
were collected in August 2013 and March 2014 towards Orion KL
(see \citealt{Tercero2010} and \citealt{Lopez2014}, for information
about the previous data set).
Frequencies in the ranges 80.7$-$116, 122.7$-$161.2, 199.7$-$291.0,
291.4$-$306.7\,GHz,
were observed with the EMIR receivers connected to
the FFTS (200\,kHz of spectral resolution) spectrometers.
We pointed towards IRc2 source at $\alpha$$_{2000.0}$\,=\,5$^h$35$^m$14$\secsecp$5,
$\delta$$_{2000.0}$\,=\,$-$5$^{\circ}$22$'$30$\secp$0,
corresponding to the survey position (see Sect.\,\ref{sect_dis}).
We observed an additional 
position to target the CR: $\alpha$$_{2000.0}$\,=\,5$^h$35$^m$14$\secsecp$3,
$\delta$$_{2000.0}$\,=\,$-$5$^{\circ}$22$'$37$\secp$0
(see Sect.\,\ref{sect_dis}).
The observations were performed using the wobbler
switching mode with a beam throw in azimuth of $\pm$120$''$. 
The intensity scale was calibrated using the
atmospheric transmission model (ATM, \citealt{Cernicharo1985};
\citealt{Pardo2001}). Focus and pointing were checked every
1$-$2 hours on planets or nearby quasars.
System temperatures were in the range of 100$-$800\,K from the
lowest to  highest frequencies. 
Half power beam width (HPBW) ranged from 31$''$ to 8$''$
from 80 to 307\,GHz (HPBW[arcsec]=2460/Freq.[GHz]).
The data were reduced using the GILDAS package\footnote{http://www.iram.fr/IRAMFR/GILDAS}.

\textbf{ALMA SV:} The ALMA Science Verification (SV) 
data\footnote{http://almascience.eso.org/almadata/sciver/OrionKLBand6/}
were taken in January 2012 towards the IRc2 region in
Orion. The observations were carried out with 16 antennas
of 12\,m in Band 6 (213.715-246.627\,GHz).
The primary beam was $\simeq$27$''$.
Spectral resolution was 0.488\,MHz corresponding to a velocity resolution
of 0.64\,km\,s$^{-1}$.
The observations were centred on coordinates:
$\alpha$$_{J2000}$\,=\,05$^h$35$^m$14$\secsecp$35, 
$\delta$$_{J2000}$\,=\,$-$05$^{\circ}$22$'$35$\secp$00.
The CASA software\footnote{http://casa.nrao.edu} was used for initial processing and then
the visibilities were exported to the GILDAS package.
The line maps were cleaned using the HOGBOM algorithm \citep{Hogbom1974}.
The synthesized beam ranged from 2$\secp$00$\times$1$\secp$48 with a PA of 176$^{\circ}$
at 214.0\,GHz to 1$\secp$75$\times$1$\secp$29 with a PA of 164$^{\circ}$ at 246.4\,GHz.
The brightness temperature to flux density conversion
factor is 9\,K for 1\,Jy per beam. 

\section{Results}
\subsection{Search for $trans$ Ethyl Methyl Ether}
\label{sect_res_tEME}

\begin{figure}[ht]
\begin{center}
\includegraphics[angle=0,scale=.8]{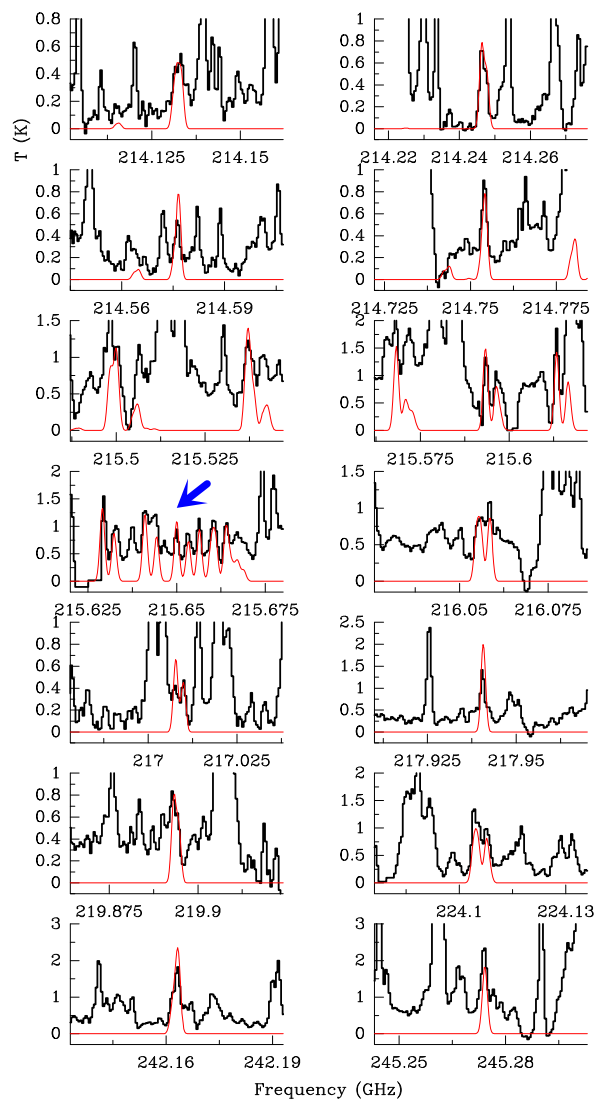}
\caption{Selected lines of $trans$ Ethyl Methyl Ether, $t$-CH$_3$CH$_2$OCH$_3$, 
towards Orion KL detected with the ALMA interferometer in Position A (see text). A $v_{LSR}$ of
+7.5\,km\,s$^{-1}$ is assumed.}
\label{fig_alma}
\end{center}
\end{figure}

\begin{figure}[ht]
\begin{center}
\includegraphics[angle=0,scale=.46]{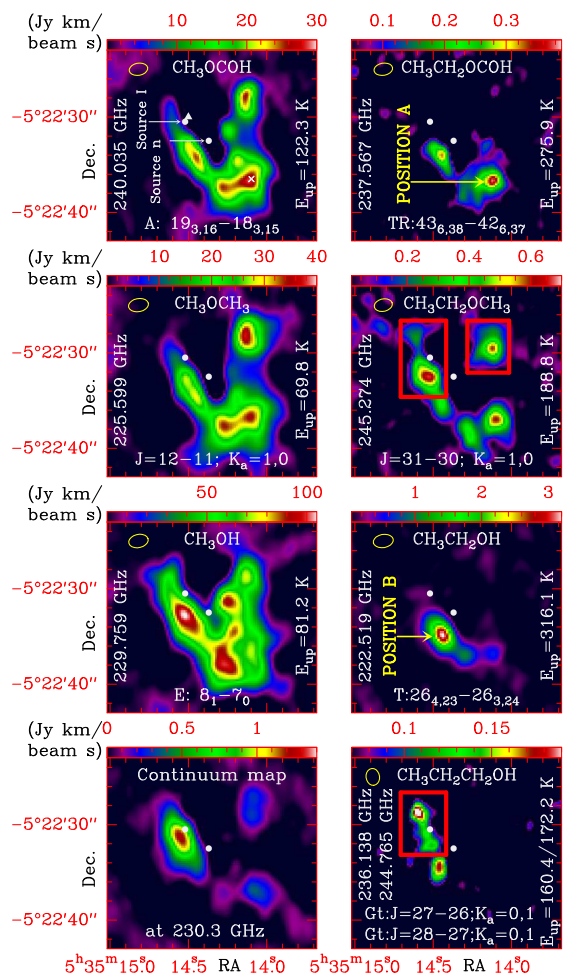}
\caption{ALMA maps of organic saturated O-bearing molecules
in Orion KL which have been detected containing both the methyl and the ethyl group, as
well as a map of Gt-n-propanol and a continuum map at the central frequencies of the ALMA SV band ($\sim$230\,GHz).
Emission that probably arises from blended species in these maps is confined inside
red rectangles
The yellow ellipse at the top left corner of the maps
represents the ALMA synthetic beam.
\textit{Triangle symbol:} IRAM 30m `survey position' (see Sect.\,\ref{sect_obs}).
\textit{Cross symbol:} IRAM 30m compact ridge position (see Sect.\,\ref{sect_obs}).
\textit{Position A:} Compact ridge (coordinates $\alpha$$_{2000.0}$\,=\,5$^h$35$^m$14$\secsecp$1,
$\delta$$_{2000.0}$\,=\,$-$5$^{\circ}$22$'$37$\secp$9).
\textit{Position B:} south hot core (coordinates $\alpha$$_{2000.0}$\,=\,5$^h$35$^m$14$\secsecp$4,
$\delta$$_{2000.0}$\,=\,$-$5$^{\circ}$22$'$34$\secp$9).}
\label{fig_mapas}
\end{center}
\end{figure}

\textbf{ALMA SV data:} Frequency predictions from \citet{Fuchs2003} and dipole moments 
measured by \citet{Hayashi1975}
of tEME were implemented in MADEX \citep{Cernicharo2012}
to model the emission of this species and search for it towards Orion KL.
Using the ALMA SV data, 
we extracted the averaged spectrum over 5$\times$5 pixels (1$''$$\times$1$''$) around
the CH$_3$OCH$_3$ emission peak of the CR component (Position A; see Sect.\,\ref{sect_dis}).
The advantage
of ALMA with respect to single dish telescope data (see below)
is the drastic reduction of the confusion limit.
The ALMA SV data show the presence of tEME as shown in Fig.\,\ref{fig_alma}
(selected lines)
and Fig.\,\ref{fig_alma_todo} (all lines favourable for detection
[corresponding to $b$$-$type transitions with 
upper level energies up to 300\,K and large line strenghts, $S_{ij}$$\geq$1]
present in the ALMA SV frequency range).
The model that best fits the data is shown with the red line.
The assumed parameters 
are a source size of 3$''$,
$v$$_{LSR}$\,=\,+7.5\,km\,s$^{-1}$, $\Delta$$v$\,=\,2.0\,km\,s$^{-1}$, and $T_K$\,=\,100$\pm$20\,K. Using MADEX 
and assuming local thermodynamic equilibrium (LTE), we obtain $N_{g.s.}$(tEME)\,$\leq$\,(4.0$\pm$0.8)$\times$10$^{15}$\,cm$^{-2}$.
In our models, rotation temperature and column density values are given with their
corresponding uncertainty and we  obtained them by fitting all available lines by eye.
We adopted the source size  in agreement with the emission of the maps (see below).
In addition, a considerable number of unblended features allows us to fix the radial velocities and line widths.
According to our model, in the ALMA frequency range only 33\% of 
the detectable lines of tEME (102 lines)
are totally hidden by the emission of stronger lines of other species.
At least 46 lines (45\% of the detectable lines) shown in Fig.\,\ref{fig_alma_todo} are free of blending, i.e. these lines are present at the expected radial velocity and there are no other species
with significant intensity at the same observed frequency ($\pm$3 MHz). 
Another point to ensure this tentative detection is that the forest of lines emitted by tEME
between 215.5 and 215.7\,GHz is not covered by lines of abundant
molecules in the source allowing the detection of several lines that follow a straightened pattern
(see Fig.\,\ref{fig_alma}). Hence,
there are several clues that could reveal
the presence of this species in the CR of Orion KL,
but further
analysis exploring new available ALMA data and modelling all
the molecular content of the CR is needed to give the
definitive detection in space of tEME.
Table\,\ref{tab_1}, which is only available online, gives line parameters and blends of
all lines of favourable transitions in the ALMA SV data.
The spatial distribution of tEME is shown in Fig.\,\ref{fig_mapas}.
Lines that we found to be unblended at
the Position A appear blended with emission from other components in the averaged spectrum
(see the case of the 30m data). 
We  selected a line at 245.274\,GHz, which is
mixed with some emission from extreme velocities of $^{34}$SO$_2$ and SO$_2$. 
Nevertheless, the emission of tEME at Position A in Fig.\,\ref{fig_mapas}
is not blended (see Sect.\,\ref{sect_dis}).

\textbf{IRAM 30m data:} 
To search for tEME in the IRAM data, 
a synthetic spectra of tEME (red curve in Fig.\,\ref{fig_30m}, only available online) was
obtained with MADEX assuming LTE and
adopting the following physical parameters:
source diameter 3$''$, $T_K$\,=\,100$\pm$30\,K, $v_{LSR}$\,=\,+7.5\,km\,s$^{-1}$, $\Delta$$v$\,=\,1.5\,km\,s$^{-1}$;
and a column density of (9$\pm$3)$\times$10$^{15}$\,cm$^{-2}$ for the ground state (g.s.) of tEME.
According to our model,
all favourable lines for detection in the 30m data were detected or
were
blended with features from more abundant species.
Nevertheless, owing to the weakness of the
features ($T_{MB}$\,$<$0.1\,K at 3\,mm, $T_{MB}$\,$<$\,0.2\,K
at 2\,mm, and  $T_{MB}$\,$<$1\,K at 1.3$-$0.9\,mm) and the high
level of line confusion at $\sim$\,1\,mm, only a few lines were mostly free
of blending with other species in this domain.
Whereas the synthetic beam of the ALMA SV 
is 1$\secp$90$\times$1$\secp$40
in the 30m the beam diameter ranging from 30$''$ to 8$''$.
 Therefore, in the 30m data,
the spectrum is a mix of all molecules from all source components
(average spectrum over the beam) given rise to a high level of line blending and line confusion.
Table\,\ref{tab_2}, which is only available online, shows
line parameters, intensity provided by the model, and blends of
all lines of favourable transitions in the 30m data.


\subsection{Search for $gauche$-$trans$-n-propanol}
\label{sect_res_prop}

All lines of Gt-CH$_3$CH$_2$CH$_2$OH, an isomer of C$_3$H$_8$O (as well as tEME),
reported by \citet{Maeda2006} and
the dipole moments 
from \citet{Abdurakhmanov1969} were used to
derive its rotational constants and to implement this species in MADEX.
We conducted the search for Gt-n-propanol in the ALMA SV data  at two different positions:
Position A
and the position where the emission peak of ethanol is located (Position B; see Sect.\,\ref{sect_dis}).
We assign several unidentified lines in the source at Position B to this species.
According to our model ($d_{sou}$\,=\,3$''$, $v$$_{LSR}$\,=\,+8.0\,km\,s$^{-1}$, $\Delta$$v$\,=\,3.0\,km\,s$^{-1}$, $T_K$\,=\,100$\pm$20\,K,
and $N_{g.s}$\,$\leq$\,(1.0$\pm$0.2)$\times$10$^{15}$\,cm$^{-2}$), many of the lines are below the detection limit although the strongest features are detected.
Unfortunately, several lines remain blended (see Fig.\,\ref{fig_alma_propanol}, only available online).
A few lines of this species are also detected in the IRAM 30m data at the survey position (Fig.\,\ref{fig_30m} bottom panel, which is only available online; 
model parameters: $d_{sou}$\,=\,3$''$, $v$$_{LSR}$\,=\,+8.0\,km\,s$^{-1}$, $\Delta$$v$\,=\,1.5\,km\,s$^{-1}$, $T_K$\,=\,100$\pm$20\,K,
and $N_{g.s}$\,$\leq$\,(2.0$\pm$0.4$)\times$10$^{15}$\,cm$^{-2}$).
Table\,\ref{tab_3}, which is only available online, shows line parameters
for the detected lines. 
The derived upper limit to its column density (assuming the same physical parameters than those of the tEME ALMA model)
at Position A is $\leq$(3.0$\pm$0.6)$\times$10$^{14}$\,cm$^{-2}$.
The spatial distribution of this species around Position B is shown in Fig.\,\ref{fig_mapas}.
To perform the ALMA map, we averaged the emission between $v$$_{LSR}$ 6 and 9\,km\,s$^{-1}$
of two lines (lines at 236.138 and 244.765\,GHz). Emission around source $I$
should be due to other less abundant species in Orion (we did not find
Gt-n-propanol at these positions).

\section{Discussion}
\label{sect_dis}

Species containing the functional groups formate, alcohol, and ether have
been detected in Orion with both the methyl and ethyl groups
(methyl formate (MF), ethyl formate (EF), methanol, ethanol,
dimethyl ether (DME), and tEME). ALMA maps for the spatial distribution
of these species as well as Gt-n-propanol are shown in Fig.\,\ref{fig_mapas}.
To address the flux filtered out by ALMA and the accuracy of the
maps in a larger energy range,
the following discussion is also based on the maps shown in Fig.\,5 of
Feng et al. (2015; maps performed mixing SMA and IRAM 30m data) with MF, DME, methanol, and ethanol.
For MF, DME, and methanol the spatial distribution and the position of the emission
peaks are in agreement with those of the maps presented in this work (note, however,
that the ALMA maps provide a more detailed structure at small scales, i.e. $\leq$5$''$).
For ethanol, we note a more extended spatial distribution in the map of \citet{Feng2015}
mostly due to the lower energy of the transition involved. Nevertheless,
the emission peak of ethanol is located at the same position.

For the methyl species,
we note: i) a rather similar spatial structure: the three species present
the V shape distribution of several clumps (at least six) studied by \citet{Favre2011} for
the distribution of MF, which was mapped using data from the Plateau de Bure Interferometer (PdBI);
ii) that although \citet{Brouillet2013} probed a striking similarity between the spatial distributions of CH$_3$OCH$_3$
and CH$_3$OCOH, we found some differences in the relative intensities of both species.
These differences
could be mostly due to different excitation temperatures of the involved transitions;
and iii) although methanol also follows this V shape structure, a displacement of the
intensity peaks is observed with respect to MF. 
This behaviour suggests methanol as a possible precursor of MF and
DME (see also \citealt{Neill2011}).

Comparing the methyl and ethyl species, we note:
i) a 
reduced spatial distribution of the three ethyl species with
respect to their methyl counterpart;
ii) the two emission peaks of EF are correlated with those
found in MF;
iii) the emission peak of tEME is at the same position as
the DME peak at the CR (Position A);
and iv) the emission peak of ethanol (Position B) is displaced 2$''$ south-west from the
methanol peak.

Concerning the ethyl and propyl species, we note:
i) a close correlation between EF and tEME;
and ii) ethanol also presents  a "V" shape structure (see Fig.\,5 of
\citealt{Feng2015})  with the bulk of the emission located
away from the CR and coinciding with that of Gt-n-propanol.
The ethanol/propanol peak is displaced 1$\secp$5 south from the ethylene glycol 
(CH$_2$OH)$_2$
peak \citep{Brouillet2015}, which is a double alcohol and we could naively expect
to have the same spatial distribution.
Whereas the ethylene glycol peak corresponds to the $^{13}$CH$_3$OH peak,
the ethanol/propanol peak is the same as that of deuterated methanol (CH$_2$DOH; see \citealt{Peng2012}).


%

\begin{table}
\begin{center}
\caption{Column densities and ratios} \label{tab_cd}
\scalebox{.83}{
\begin{tabular}{lll}
\hline\hline 
  Species     & $N$$_{g.s.}$ ($\times$10$^{15}$) [cm$^{-2}$] & $N$ Ratio \\
\hline  
  CH$_3$OCH$_3$ (DME)             & 600$\pm$120$^{(a),(b)}$ & \\
  CH$_3$CH$_2$OCH$_3$ (tEME)      & $\leq$4.0$\pm$0.8$^{(a)}$ & DME/tEME\,$\geq$\,150\\
        CH$_3$CH$_2$OCH$_2$CH$_3$       & $\leq$1.0$\pm$0.2$^{(a)}$ &  DME/Tt-DEE\,$\geq$\,600\\
  (Tt-DEE)$\dagger$               &                            & tEME/Tt-DEE\,$\geq$\,4\\  
        CH$_3$OCHCH$_2$                 & $\leq$0.5$\pm$0.1$^{(a)}$ & DME/$cis$-MVE\,$\geq$\,1200\\
        ($cis$-MVE)$\dagger$$\dagger$   &                           &  tEME/$cis$-MVE\,$\geq$\,9\\
        CH$_3$OCOH (MF)              & 240$\pm$50$^{(a),(b),(c)}$ &\\
        CH$_3$CH$_2$OCOH (EF)        & 2.0$\pm$0.4$^{(a),(d)}$ & MF/EF\,$\simeq$\,120\\
        CH$_3$OH (MetOH)             & 2700$\pm$500$^{(b),(e),(f)}$ & \\
        CH$_3$CH$_2$OH (EtOH)        & 60$\pm$10$^{(b),(d),(e)}$ & MetOH/EtOH\,$\simeq$\,45\\
        Gt-CH$_3$CH$_2$CH$_2$OH      & 1.0$\pm$0.2$^{(e)}$ & MetOH/PropOH\,$\simeq$\,2700\\
         (PropOH)                    &                 & EtOH/PropOH\,$\simeq$\,60\\
\hline  
\end{tabular}
}
\end{center}
{\footnotesize{$\dagger$: $trans$-$trans$ Diethyl ether. $\dagger$$\dagger$: $cis$ Methyl vinyl ether.
\textit{\textbf{(a):}} Position A; same physical parameters of the ALMA tEME model
(see Sect.\,\ref{sect_res_tEME}). \textit{\textbf{(b):}} Three kinetic temperatures: 
50$\pm$10, 150$\pm$30, and 250$\pm$75\,K. \textit{\textbf{(c):}} b type lines fitted 
(a type lines are optically thick); another component has been
included to properly fit the observed line profiles ($v_{LSR}$\,=\,+9\,km\,s$^{-1}$,
$\Delta$$v$\,=\,4\,km\,s$^{-1}$, $T_K$\,=\,150$\pm$30\,K, 
$N_{g.s}$\,=\,(1.0$\pm$0.2$)\times$10$^{17}$\,cm$^{-2}$). \textit{\textbf{(d):}} $trans$+$gauche$. 
\textit{\textbf{(e):}} Position B; assuming the same physical parameters of the ALMA Gt-n-propanol model
(see Sect.\,\ref{sect_res_prop}). \textit{\textbf{(f):}} $^{12}$C/$^{13}$C\,=\,45 \citep{Tercero2010}.                 
}}
\end{table}

Table\,\ref{tab_cd} shows derived column densities and ratios
for related species.
The derived ratios and the spatial distribution of these molecules
suggest important gas phase processes after the evaporation of the mantles of dust
grains in hot cores. Possible reactions of the methoxy radical (CH$_3$O),
detected recently in space \citep{Cernicharo2012b}, with other species could lead to the increase 
of chemical complexity in hot cores and hot corinos \citep{Balucani2015}. The spatial
stratification of the different species also suggests the time dependent effects on the chemistry
of the gas. The detection of the less stable isomers of some species \citep{Tercero2013}
also points in this direction. 

To summarize,
a combined IRAM 30m and ALMA SV data study allows us to provide a solid
starting point to assess the identification of
tEME in the ISM. In addition, some unidentified lines in the source have been assigned to another C$_3$H$_8$O isomer, Gt-n-propanol. 
ALMA maps show different spatial distributions for these species.
Whereas tEME seems to mainly arises from the CR component (as well as EF) [Position A], emission from Gt-n-propanol
could be located at the south hot core 
(at the same position as the emission peak of ethanol) [Position B].
The CR is no longer the main host of all organic saturated O-bearing species in Orion
(see also \citealt{Peng2013} for the spatial distribution of acetone and A. L\'opez et al. in preparation for
the acetic acid emission).

\begin{acknowledgements}
We thank Marcelino Ag\'undez
for carefully reading the paper and providing useful comments and suggestions.
B.T., J.C., and A.L. thank the Spanish MINECO for funding support under grants CSD2009-00038, 
AYA2009-07304, and AYA2012-32032 and also the ERC for funding support under grant 
ERC-2013-Syg-610256-NANOCOSMOS.
\end{acknowledgements}

\Online

\begin{appendix}
\section{Online Figures and Tables}

\begin{figure*}
\includegraphics[angle=0,scale=.9]{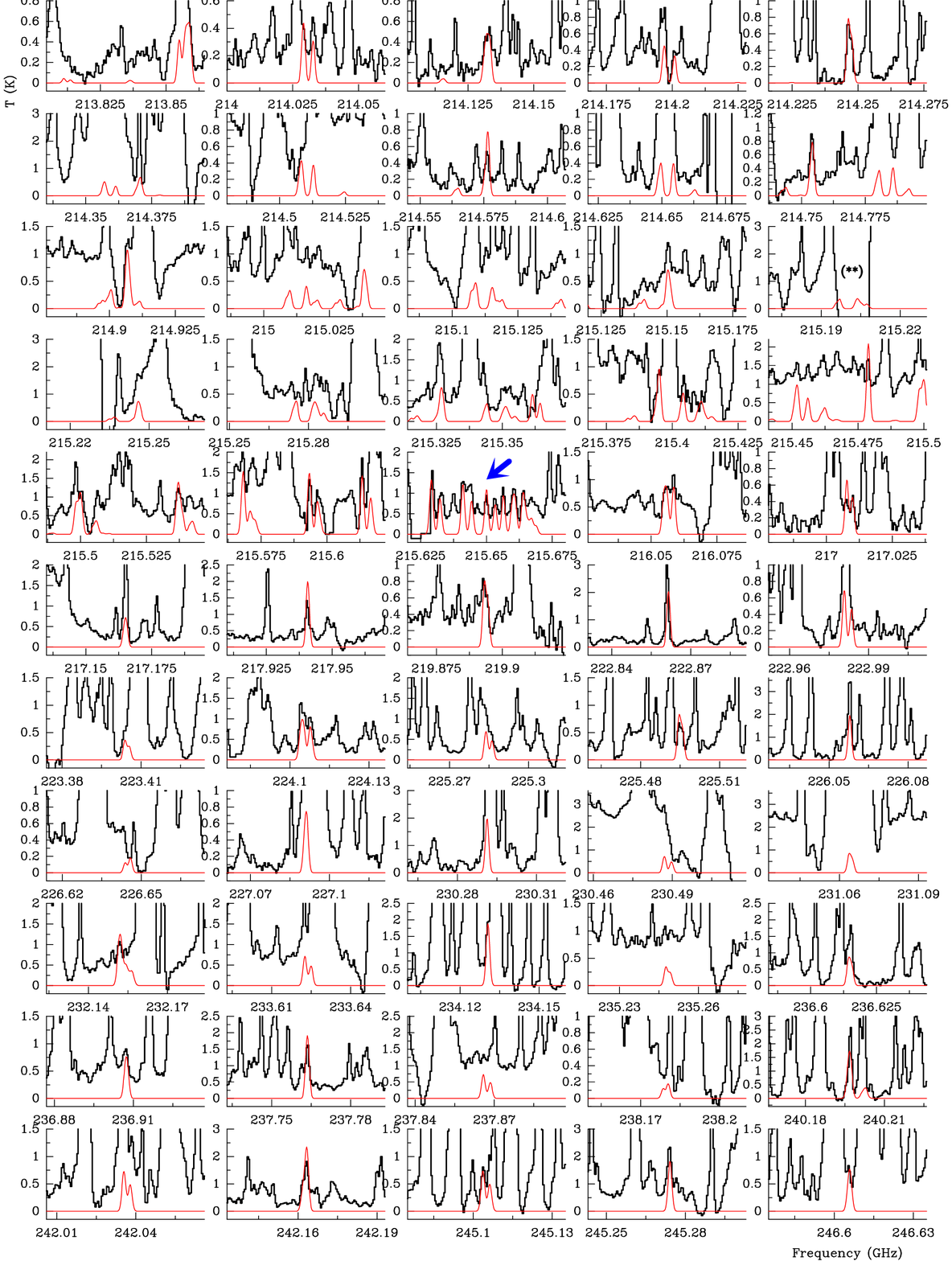}
\caption{Lines of $trans$ Ethyl Methyl Ether, $t$-CH$_3$CH$_2$OCH$_3$, 
towards Orion KL detected with the ALMA interferometer in Position A (see text).
(**): Features blended with SO (see Table A.1; artifacts in the spectrum due to the cleaning process). A $v_{LSR}$ of
+7.5\,km\,s$^{-1}$ is assumed.}
\label{fig_alma_todo}
\end{figure*}

\begin{figure*}[ht]
\includegraphics[angle=0,scale=.9]{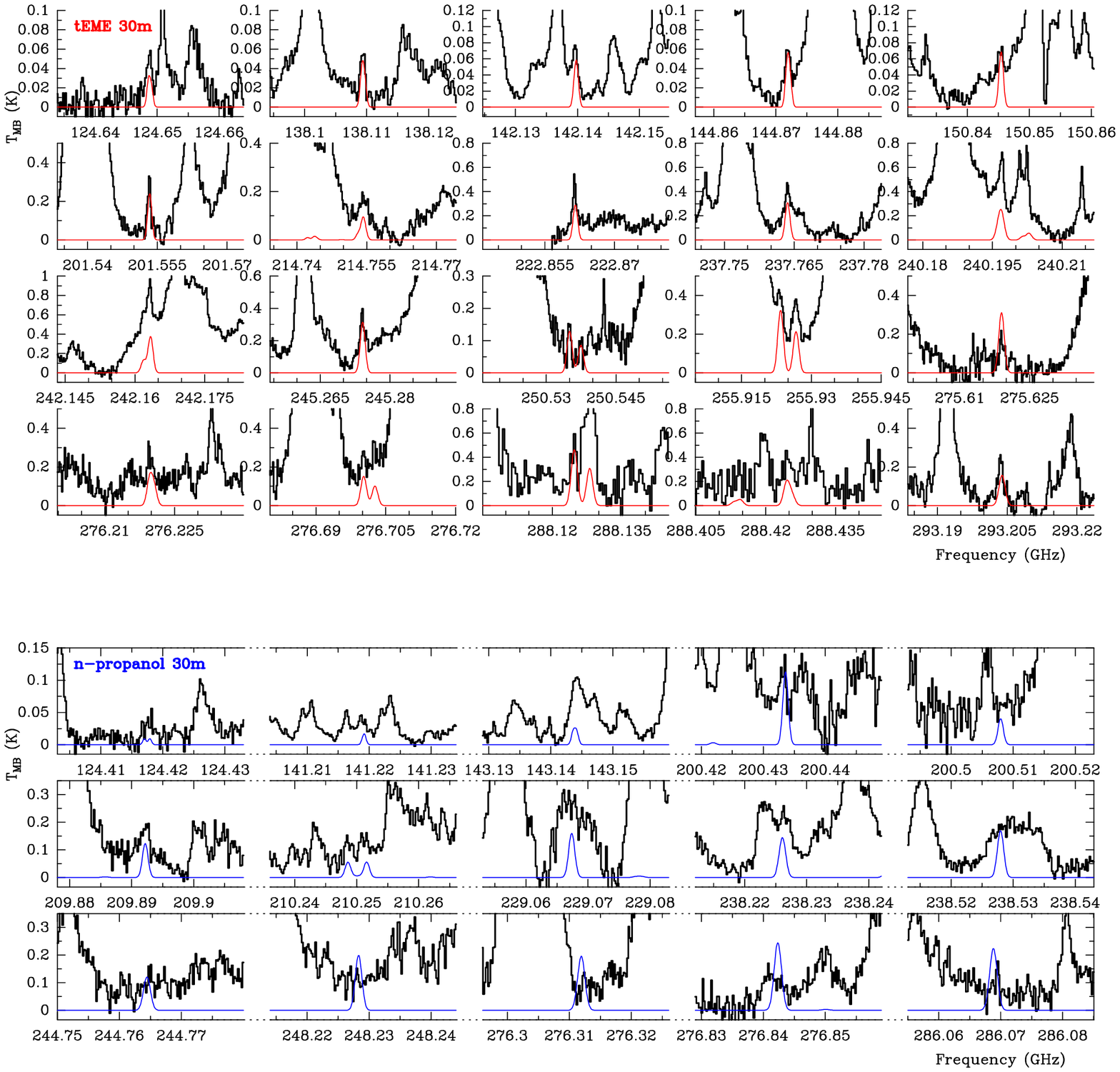}
\caption{\textit{Top panel}: selected lines of $trans$ Ethyl Methyl Ether, $t$-CH$_3$CH$_2$OCH$_3$, 
towards Orion KL detected with the IRAM 30m telescope. Data in the frequency range 
124$-$151\,GHz are those of the survey position. From 201 to 293.5\,GHz
the data are those of the CR (see Sect.\,\ref{sect_obs}), where the emission peak of organic
saturated O-rich species such as dimethyl
ether (CH$_3$OCH$_3$) and methyl formate (CH$_3$OCOH) is located \citep{Favre2011,Brouillet2013}. A $v_{LSR}$ of
+7.5\,km\,s$^{-1}$ is assumed. \textit{Bottom panel}: selected lines of $gauche$-$trans$-n-Propanol, Gt-n-CH$_3$CH$_2$CH$_2$OH, 
towards Orion KL detected with the IRAM 30m telescope. A $v_{LSR}$ of
+7.5\,km\,s$^{-1}$ is assumed.}
\label{fig_30m}
\end{figure*}

\begin{figure*}
\includegraphics[angle=0,scale=.9]{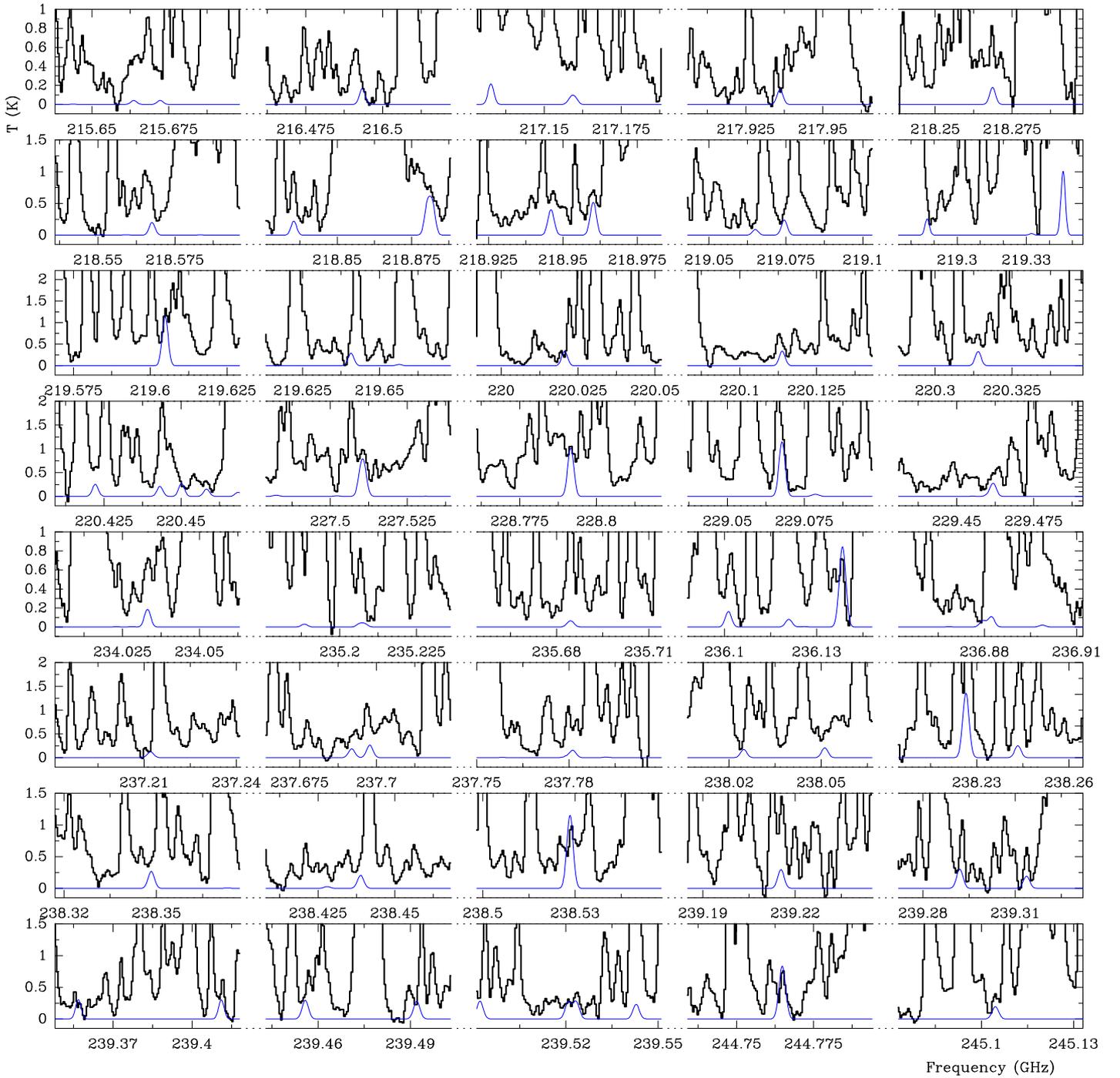}
\caption{Lines of $gauche$-$trans$-n-propanol, Gt-n-CH$_3$CH$_2$CH$_2$OH, 
towards Orion KL detected with the ALMA interferometer in Position B (see text). A $v_{LSR}$ of
+8\,km\,s$^{-1}$ is assumed.}
\label{fig_alma_propanol}
\end{figure*}

\clearpage

\onecolumn
\scriptsize

\begin{longtable}{lccrrccc}
\caption{Lines of $trans$-CH$_3$CH$_2$OCH$_3$ in ALMA SV data} \label{tab_1}\\

\hline\hline
  Species       & Transition     & Predicted  & $E_{upp}$ & $S_{ij}$ & Observed & $T$ & Blends \\
                   & $J_{K_a,K_c}-J'_{K'_a,K'_c}$ &  frequency (MHz)  &    (K)      &   &   frequency (MHz)   &    (K)& \\
\hline
\endfirsthead
\caption{continued.}\\
\hline\hline
  Species       & Transition     & Predicted  & $E_{upp}$ & $S_{ij}$ & Observed & $T$ & Blends \\
                   & $J_{K_a,K_c}-J'_{K'_a,K'_c}$ &  frequency (MHz)  &    (K)      &   &   frequency (MHz)   &    (K)& \\
\hline
\endhead
\hline
\endfoot

tEME-EE'  &   31$_{5,27}$-31$_{4,28}$      &  213854.674 & 220.5 & 16.34 &    ...               & ...  &  CH$_3$CH$_2$OH\\
tEME-EE   &   31$_{5,27}$-31$_{4,28}$      &  213854.982 & 220.5 & 16.44 &    ...                  & ...  &  "\\
tEME-AE   &   31$_{5,27}$-31$_{4,28}$      &  213855.220 & 220.5 & 16.40 &    ...                  & ...  &  "\\
tEME-EE   &   28$_{5,23}$-28$_{4,24}$      &  213856.708 & 185.7 & 14.17 &    ...                  & ...  &  "\\
tEME-AE   &   28$_{5,23}$-28$_{4,24}$      &  213857.473 & 185.7 & 13.94 &    ...                  & ...  &  "\\
tEME-EE'  &   28$_{5,23}$-28$_{4,24}$      &  213857.509 & 185.7 & 13.69 &    ...                  & ...  &  "\\
tEME-EA   &   31$_{5,27}$-31$_{4,28}$      &  213858.239 & 220.5 & 16.52 &    ...                  & ...  &  "\\
tEME-AA   &   31$_{5,27}$-31$_{4,28}$      &  213858.632 & 220.5 & 16.53 &    ...                 & ...  &  "\\
\end{longtable}

\flushleft
{\footnotesize{
{\bf{Note.-}} Lines of $trans$-CH$_3$CH$_2$OCH$_3$ (tEME) ground state 
present in the spectral scan of Orion KL from the 
ALMA interferometer. Column 1 indicates the species, Column 2 gives the 
transition, Column 3 the predicted frequency, Column 4 upper level energy, 
Column 5 the line strength, Column 6 observed frequency at the peak channel of the line (relative to a $v_{LSR}$ of +7.5\,km\,s$^{-1}$), 
Column 7 brightness temperature at the peak channel of the line,
and Column 8 shows blends with other molecular species.\\
$\dagger$ Blended with previous line.      
}}

\clearpage

\onecolumn
\scriptsize

\begin{longtable}{lccrrcccc}
\caption{Lines of $trans$-CH$_3$CH$_2$OCH$_3$ in 30m data} \label{tab_2}\\

\hline\hline
  Species       & Transition     & Predicted  & $E_{upp}$ & $S_{ij}$ & Observed$^{(1)}$ & Observed$^{(1)}$ & Model$^{(2)}$ & Blends \\
                   & $J_{K_a,K_c}-J'_{K'_a,K'_c}$ &  frequency (MHz)  &    (K)      &   &   frequency (MHz)   & $T_{MB}$ (K)& $T_{MB}$ (K) & \\
\hline
\endfirsthead
\caption{continued.}\\
\hline\hline
  Species       & Transition     & Predicted  & $E_{upp}$ & $S_{ij}$ & Observed$^{(1)}$ & Observed$^{(1)}$ & Model$^{(2)}$ & Blends \\
                   & $J_{K_a,K_c}-J'_{K'_a,K'_c}$ &  frequency (MHz)  &    (K)      &   &   frequency (MHz)   & $T_{MB}$ (K)& $T_{MB}$ (K) & \\
\hline
\endhead
\hline
\endfoot
tEME-AA   &   13$_{0,13}$-12$_{1,12}$      &   88665.592 &  35.0 &  7.66 &     88665.8         &  0.05 & 0.01 & U-line\\
tEME-EA   &   13$_{0,13}$-12$_{1,12}$      &   88665.628 &  35.0 &  7.66 &            $\dagger$&       & &\\
tEME-AE   &   13$_{0,13}$-12$_{1,12}$      &   88666.042 &  35.0 &  7.66 &            $\dagger$&       & &\\
tEME-EE   &   13$_{0,13}$-12$_{1,12}$      &   88666.078 &  35.0 &  7.66 &            $\dagger$&       & &\\
tEME-EE'  &   13$_{0,13}$-12$_{1,12}$      &   88666.078 &  35.0 &  7.66 &            $\dagger$&       & &\\
tEME-EE'  &   10$_{1,10}$-9$_{0,9}$        &   97575.502 &  22.0 &  6.04 &    ...          & ...   & 0.01 & CH$_3$OH\\
tEME-EE   &   10$_{1,10}$-9$_{0,9}$        &   97575.502 &  22.0 &  6.04 &    $\dagger$        &      & &\\
tEME-AE   &   10$_{1,10}$-9$_{0,9}$        &   97575.556 &  22.0 &  6.04 &    $\dagger$        &      & &\\
\end{longtable}

\flushleft
{\footnotesize{
{\bf{Note.-}} Lines of $trans$-CH$_3$CH$_2$OCH$_3$ (tEME) ground state 
present in the spectral scan of Orion KL from the 
30m telescope. Column 1 indicates the species, Column 2 gives the 
transition, Column 3 the predicted frequency, Column 4 upper level energy, 
Column 5 the line strength, Column 6 observed frequency at the peak channel of the line (relative to a $v_{LSR}$ of +7.5\,km\,s$^{-1}$), 
Col. 7 main beam temperature at the peak channel of the line,
and Column 8 shows blends with other molecular species.\\
(1) Observed frequencies and intensities are not provided for features
that appear totally blended with lines from other species.
(2) We address all features provided by our model with $T_{MB}$\,$>$\,0.01\,K,
$T_{MB}$\,$>$\,0.02\,K, and $T_{MB}$\,$>$\,0.03\,K in the frequency ranges between
80.7$-$116, 122.7$-$150, and 150$-$306.7\,GHz, respectively.
$\dagger$ Blended with previous line.   
}}

\clearpage

\onecolumn
\scriptsize

\begin{longtable}{lccrrccc}
\caption{Detected lines of $gauche$-$trans$-n-CH$_3$CH$_2$CH$_2$OH.} \label{tab_3}\\

\hline\hline
  Species       & Transition     & Predicted  & $E_{upp}$ & $S_{ij}$ & Observed & $T$ & Blends/ \\
                   & $J_{K_a,K_c}-J'_{K'_a,K'_c}$ &  frequency (MHz)  &    (K)      &   &   frequency (MHz)   &    (K)& Comments\\
\hline
\endfirsthead
\caption{continued.}\\
\hline\hline
  Species       & Transition     & Predicted  & $E_{upp}$ & $S_{ij}$ & Observed & $T$ & Blends/ \\
                   & $J_{K_a,K_c}-J'_{K'_a,K'_c}$ &  frequency (MHz)  &    (K)      &   &   frequency (MHz)   &    (K)& Comments \\
\hline
\endhead
\hline
\endfoot

Gt-n-propanol&   12$_{7,6}$-12$_{6,6}$       &  124417.241 &  58.0 &  4.25 &   124417.1     &  0.04 &30m; U-line\\ 
Gt-n-propanol&   12$_{7,5}$-12$_{6,6}$       &  124417.256 &  58.0 &  4.89 &       $\dagger$&      &\\
Gt-n-propanol&   12$_{7,6}$-12$_{6,7}$       &  124418.125 &  58.0 &  4.89 &  124418.2$\dagger$&  0.06 & 30m; U-line\\
Gt-n-propanol&   12$_{7,5}$-12$_{6,7}$       &  124418.139 &  58.0 &  4.25 &       $\dagger$&      &\\
Gt-n-propanol&   15$_{2,14}$-14$_{1,13}$     &  141219.413 &  55.4 &  9.14 &   141219.8     &  0.04 &30m\\ 
Gt-n-propanol&    6$_{5,2}$-5$_{4,1}$        &  143143.854 &  21.1 &  4.51 &   143144.6     &  0.10 &30m; CH$_3$CH$_2$CN $\varv_{12}$=1\\ 
Gt-n-propanol&    6$_{5,1}$-5$_{4,1}$        &  143143.873 &  21.1 &  4.65 &       $\dagger$&      &\\
Gt-n-propanol&    6$_{5,2}$-5$_{4,2}$        &  143144.406 &  21.1 &  4.65 &       $\dagger$&      &\\
\end{longtable}

\flushleft
{\footnotesize{
{\bf{Note.-}} Lines of 
$gauche$-$trans$-n-CH$_3$CH$_2$CH$_2$OH (Gt-n-propanol) ground state 
present in the spectral scan of Orion KL from the IRAM-30 m telescope and the 
ALMA interferometer. Column 1 indicates the species, Column 2 gives the 
transition, Column 3 the predicted frequency, Column 4 upper level energy, 
Column 5 the line strength, Column 6 observed frequency at the peak channel of the line (relative to a $v_{LSR}$ of 
+8.0\,km\,s$^{-1}$), 
Col. 7 temperature at the peak channel of the line (main beam temperature for the IRAM data),
and Column 8 shows blends with other molecular species and comments.\\
$\dagger$ Blended with previous line.
}}

\end{appendix}

\end{document}